\documentclass[pre,english,showpacs,twocolumn,superscriptaddress]{revtex4-1}

\usepackage[english]{babel}
\usepackage[dvips]{epsfig}
\usepackage{amssymb}
\usepackage{amsmath}
\usepackage{amsbsy}
\usepackage{dcolumn}

\usepackage{color}

\newcommand{\vscb} {v_{\mathrm{SCB}}}
\newcommand{\vdws} {v_{\mathrm{DWS}}}
\newcommand{\zroi} {z_{\mathrm{ROI}}}
\newcommand{\um} {~\mu\mathrm{m}}
\newcommand{\mm} {~\mathrm{mm}}
\newcommand{\epsdot} {\dot{\varepsilon}}
\definecolor {myc} {rgb} {0,0,0}  

\usepackage{lineno,hyperref} 

\begin{document}


\title{Space-resolved diffusing wave spectroscopy measurements of the macroscopic deformation and the microscopic dynamics in tensile strain tests}

\author{Med-Yassine Nagazi}
\affiliation{Laboratoire Charles Coulomb (L2C), UMR 5221 CNRS-Universit\'{e} de Montpellier,
Montpellier, France}
\affiliation{Formulaction, L'Union, F-France}

\author{G\'{e}rard Meunier}
\affiliation{Formulaction, L'Union, F-France}

\author{Philippe Marguer\`{e}s}
\affiliation{Institut Cl\'{e}ment Ader (ICA), UMR 5312 CNRS-Universit\'e F\'ed\'erale Toulouse Midi-Pyr\'en\'ees, Toulouse, F-France}

\author{Jean-No\"{e}l P\'{e}ri\'{e}}
\affiliation{Institut Cl\'{e}ment Ader (ICA), UMR 5312 CNRS-Universit\'e F\'ed\'erale Toulouse Midi-Pyr\'en\'ees, Toulouse, F-France}

\author{Luca Cipelletti}
\affiliation{Laboratoire Charles Coulomb (L2C), UMR 5221 CNRS-Universit\'{e} de Montpellier,
Montpellier, France}

\begin{abstract}
We couple a laser-based, space-resolved dynamic light scattering apparatus to a universal traction machine for mechanical extensional tests. We perform simultaneous optical and mechanical measurements on polyether ether ketone, a semi-crystalline polymer widely used in the industry. Due to the high turbidity of the sample, light is multiply scattered by the sample and the diffusing wave spectroscopy (DWS) formalism is used to interpret the data. Space-resolved DWS yields spatial maps of the sample strain and of the microscopic dynamics. An excellent agreement is found between the strain maps thus obtained and those measured by a conventional stereo-digital image correlation technique. The microscopic dynamics reveals both affine motion and plastic rearrangements. Thanks to the extreme sensitivity of DWS to displacements as small as 1 nm, plastic activity and its spatial localization can be detected at an early stage of the sample strain, making the technique presented here a valuable complement to existing material characterization methods.
\end{abstract}

\maketitle




\section{Introduction}

The characterization of the mechanical properties of a material is a key step in the development of new materials, as well as in tests of existing materials for industrial applications and in quality control programs, e.g. during production. Mechanical tests typically consist in imposing a strain to the sample while measuring the resistance force exerted by it, or, alternatively, in measuring the strain for a given applied force. This is the case, e.g., of tensile tests, where a universal traction machine (UTM) imposes a tensile strain and measures the associated tensile force. These tests yield basic information on both the linear elasticity (e.g. via the Young modulus $E$) and the sample behavior beyond the linear regime, up to mechanical failure. The latter may occur due to a sudden breakage, as in the fracture of fragile materials, or may stem from a very large, irreversible plastic deformation, as in ductile materials. More insight on the material behavior can be gained by coupling these macroscopic measurements to more local ones. A popular example is the measurement of the full, mesoscopic strain field, rather than the global, macroscopic strain alone. This can be achieved using stereo-digital image correlation methods~\cite{sutton_image_2009,ye_kinematic_2015}. A stereo-correlation bench (SCB) consisting in a pair of cameras takes pictures of the sample surface during the mechanical test. The surface is treated in such a way to create a contrasted pattern, e.g. by spray-painting it with a dotted pattern. Digital Imaging Correlation (DIC) techniques are subsequently used to compute the strain field by following the displacement of regions of interests (ROIs)~\cite{sutton_image_2009,ye_kinematic_2015,pan_twodimensional_2009}. By examining the strain maps valuable information can be obtained, e.g. by detecting small deviations from the uniform strain field expected in the linear regime, which may reveal at an early stage strain localization that eventually will result in macroscopic damage, such as necking.

Spray-painting, however, comes with some limitations: for some materials, it may be difficult to select a suitable paint coat~\cite{el-hajjar_adhesive_2011}. Furthermore, the paint film may partially detach from its substrate, especially for large strains~\cite{grytten_use_2009,le_cam_review_2012}, leading to artifacts in the measurement of the actual strain field. An appealing alternative is to illuminate the sample with laser light. Due to the coherence of the laser source, light backscattered or back-reflected by the sample forms a random interference pattern, termed speckle pattern~\cite{goodman_speckle_2007}. The speckle pattern has a contrasted, grainy appearance and the speckle size can be easily tuned by changing the aperture stop of the objective lens used to image the sample. Any displacement of the sample is mirrored by a displacement of the speckle pattern, so that the strain field may be reconstructed by using the same techniques as in conventional image correlations methods. However, this speckle pattern is extremely sensitive not only to a rigid displacement of the sample, but also to the relative motion of its constituents, on length scales that for turbid, multiply scattering materials can be as small as a fraction of nm~\cite{weitz_diffusing-wave_1993}. As a consequence, contrary to the pattern obtained by spray painting, not only does the laser speckle pattern of a strained sample deform, but it also decorrelates.

While the decorrelation of the speckles in a sample under stress limits the possibility of following its macroscopic strain~\cite{pan_twodimensional_2009}, it also provides an opportunity for characterizing the deformation field at the microscopic level. This has been recognized since the early developments of diffusing wave spectroscopy (DWS), a scattering technique in the multiple scattering regime~\cite{weitz_diffusing-wave_1993}. In DWS, the decorrelation of the speckle pattern generated by a turbid sample illuminated by a laser beam can be quantitatively related to the relative motion of the objects responsible for the scattering. The latter may be, e.g., tracer particles in a fluid or embedded in a solid matrix, or, more generally, the fluctuations of the refractive index associated with variations of the local density or composition of the sample. While most studies have been devoted to the case where the microscopic dynamics is induced by thermal energy, as for Brownian motion, the case where the dynamics result predominantly from an applied strain or stress has also been investigated. Wu and coworkers have related the speckle decorrelation measured by DWS to the shear rate in a fluid under Poiseuille laminar flow~\cite{wu_diffusing-wave_1990}. Bicout et al. have generalized the approach of Ref.~\cite{wu_diffusing-wave_1990} to an arbitrary deformation~\cite{bicout_diffusing_1993}, expressing the speckle patter decorrelation as a function of the strain tensor. Their formalism has been used to analyze experimental speckle decorrelation data for a random close packing of sub-millimeter glass spheres subject to thermal expansion~\cite{crassous_diffusive_2007}. This experiment has highlighted the extreme sensitivity of the technique, since relative dilations as small as $10^{-6}$ could be measured.

In Refs.~\cite{wu_diffusing-wave_1990,bicout_diffusing_1993,crassous_diffusive_2007} the speckle field was analyzed in the far field, such that the detector was illuminated by light issued from the whole scattering volume, with no spatial selectivity. In this configuration, the evolution of the speckle pattern is only due to the relative motion of the scatterers. A different approach has been introduced in a subsequent series of experiments by the group of J. Crassous~\cite{erpelding_diffusive_2008,erpelding_diffusing-wave_2013}, where the speckle patter was collected in an imaging geometry, such that different areas of the 2D detector corresponded to different regions of the sample, a slab of a solid material loaded on one side with a blade in a plane stress configuration. {\color {myc} The measurements of~\cite{erpelding_diffusive_2008,erpelding_diffusing-wave_2013} were performed well within the linear regime, for extremely small strains of order $10^{-3}$ to $10^{-5}$, such that the sample was essentially immobile in the laboratory frame of reference and the strain field could be considered to be purely affine.}

{\color {myc} In this work, we apply space-resolved DWS to investigate the behavior of semi-crystalline and amorphous polymers during tensile stress tests. In contrast to Refs.~\cite{erpelding_diffusive_2008,erpelding_diffusing-wave_2013}, we explore both the linear and non-linear regime, with strains up to about 10\%. This has two consequences: first, the sample is macroscopically deformed, the side held by the mobile grip of the UMT being displaced over several millimeters. We show that under these conditions PCI-DWS can be used to measure strain maps comparable to those obtained by conventional stereo-digital image correlation methods. Second, unlike in~\cite{erpelding_diffusive_2008,erpelding_diffusing-wave_2013}, the microscopic dynamics stems not only from the affine deformation associated with elastic response, but also from additional rearrangements. We show that these additional rearrangements allow one to detect plasticity at an early stage, before the onset of its macroscopic manifestation (necking).} The paper is organized as follows: in Sec.~\ref{sec:matmet} we briefly describe the setup and the sample preparation. Section~\ref{sec:analysis} deals with the data analysis methods, focussing in particular on space-resolved DWS. The experimental results are presented and discussed in Sec.~\ref{sec:results}, followed by some brief concluding remarks (Sec.~\ref{sec:conclusions}).

\section{Materials and setup}
\label{sec:matmet}

\subsection{Samples}
We perform tensile stress tests on polyether ether ketone (PEEK) samples. PEEK is a semi-crystalline thermoplastic polymer with applications in the automotive, aerospace, and chemical process industries. Samples are injection molded (PEEK grains  by Evonik AG are heated at $T = 375^\circ \mathrm{C}$) to obtain the shape required by the ISO 527-2 standard. A picture of the specimen is shown in fig.~\ref{fig:setup}c: the width of the gage section is 10 mm, that of the grip section is 20 mm, the distance between shoulders is 109 mm, the overall length 170 mm and the thickness 4 mm. The side shown in fig.~\ref{fig:setup}c has been spray-painted to obtain a random black and white pattern for measuring the sample deformation using a stereo-digital image correlation technique.
{\color {myc} For the quantitative interpretation of the microscopic dynamics, the following optical parameters have been used (see Sec.~\ref{sec:results} for more details): refractive index $n = 1.68$~\cite{wiederseiner_refractive-index_2011}, photon mean transport path $\ell^* = 61.1\um$, absorption length $\ell_a = 9~\mathrm{mm}$. To test the ability of PCI-DWS to capture the mesoscopic strain field, we also run} complementary measurements on polypropylene reinforced by glass fibers (PPG), prepared following the protocol of Ref.~\cite{etcheverry_glass_2012}. The shape of the PPG specimen is the same as that of the PEEK sample, except for the distance between shoulders and the overall length, which are 95 mm and 156 mm, respectively.

\subsection{Setup}

\begin{figure}
\includegraphics[width=0.9\columnwidth,clip]{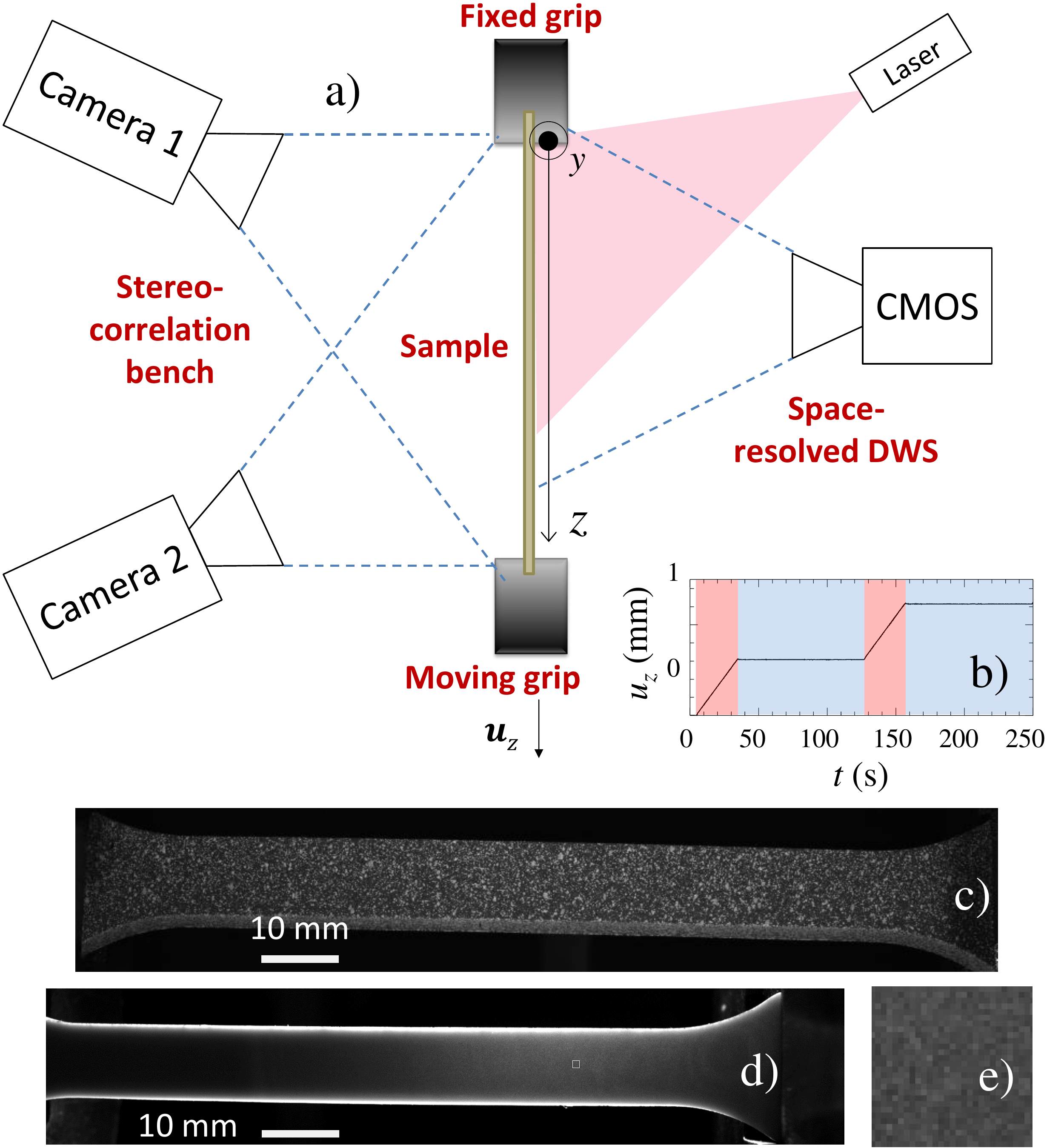}
\caption{\textbf{Experimental setup.} a): Schematic side view of the experimental apparatus. The sample is hold by two grips. The two cameras on the left are used for conventional stereo-imaging. The CMOS camera on the right images the laser light backscattered by the sample, for space-resolved DWS. b): time dependence of the displacement of the moving grip during the test, which alternates stretching phases (red shade) with relaxation phases (blue shade). For clarity, only the first 2 phases out of a total of 12 are shown. {\color {myc} c): typical image of the specimen, as captured by the SCB. d): raw speckle image for PCI-DWS. e) zoom of a small portion of the speckle image, indicated by the white square in d).}}
\label{fig:setup}
\end{figure}
Figure~\ref{fig:setup}a shows a schematic view of the setup. The sample is held in a vertical plane by {\color {myc} the two grips of the UMT (model n. 8561, by Instron); the moving (bottom) grip can be pulled downward at a controlled speed. A force sensor (maximum load: 10 kN) measures the tensile force $F_t$ exerted by the sample, which reaches at most $3.63\mathrm{~kN}$ in our experiments. The maximum displacement of the moving grip is 7.43 mm}. A conventional stereo-correlation bench (SCB) comprising CCD cameras 1 and 2 takes images of the sample surface in order to reconstruct the local displacement. Images are recorded at a rate of 0.1 Hz for subsequent processing with the stereo-DIC software. On the opposite side of the SCB, the sample is illuminated by a laser beam (Photon module by Prophotonics, with wavelength $\lambda = 635$ nm, power 20 mW, linearly polarized) in order to perform space-resolved diffusing wave spectroscopy (PCI-DWS) measurements. {\color {myc} The laser is located at 60 cm from the sample, which is illuminated under an angle of $15^{\circ}$ with respect to the normal to the sample surface}. A lens (not shown) is used to expand the laser beam so as to cover the full field of view of the CMOS camera, which images the sample surface. {\color {myc} The laser beam has an elliptical cross-section, with 1/e axes of approximately 10 cm and 2.5 cm in the sample plane.
The CMOS camera (UI-3370CP by IDS Gmbh) is located at 58 cm from the sample; it is equipped with a $2048 \times 2048$ $\mathrm{pixel}^2$ sensor, the pixel size being $5.5\times 5.5 \um^2$.} The field of view of the CMOS camera covers the full width of the sample and extends from the top grip to $z = 93.5$ mm, about 85.7\% of the sample height. {\color {myc} The camera is run at a frame rate of 10 Hz, using an exposure time of $0.995~\mathrm{ms}$. The images are recorded on a hard disk and processed off-line} as explained in Sec.~\ref{sec:SRDWS}. Stereo-DIC and PCI-DWS measurements are performed simultaneously.

The test consists of a series of 12 traction phases at a constant imposed speed of $3.33 \times 10^{-2}~\mathrm{mm~s^{-1}}$, each lasting 30 s. Between two consecutive traction phases, the sample is held at a constant strain for 90 s (relaxation phase). The graph in fig.~\ref{fig:setup}b shows the displacement $u_z$ of the lower grip as function of time $t$, for the first two traction and relaxation phases. Two sensors record both the lower grip displacement and the tensile force throughout the experiment. Self-locking grips are used, which results in a slight motion of the part of the grip in contact with the sample with respect to the part fixed to the apparatus, as seen by inspecting the  images obtained by both the SCB and the PCI-DWS cameras. In the following, we systematically correct for this effect, so that all positions and velocities are measured with respect to a reference frame where the upper part of the sample is immobile.

\section{Data analysis}
\label{sec:analysis}

\subsection{Stereo-digital correlation imaging}
An image of the sample as seen by the SCB cameras is shown in fig.~\ref{fig:setup}c. Commercial software (VIC-3D by Correlated Solutions) is used to measure 3D displacement fields on the specimen surface from images captured by the SCB~\cite{sutton_image_2009}.

The software is based on digital image correlation methods that quantify the displacement of a pattern painted on the sample surface. The analysis is performed by dividing each image in regions of interest (ROIs), whose size on the sample is $0.34 \times 0.31~\mathrm{mm}^2$. For each ROI, the SCB software provides the displacement components along three orthogonal axes; in the following we will only focus on $\Delta z$, the component of the local displacement along the pulling direction.

\subsection {Space-resolved diffusing wave spectroscopy (PCI-DWS)}
\label{sec:SRDWS}

Figures~\ref{fig:setup}d-e show images of the sample illuminated by the laser beam: the distinctive speckle pattern is due to the interference between photons backscattered by the sample and emerging at the sample surface. The PEEK and PPG samples are highly turbid, due to significant spatial fluctuations of their refractive index. Thus, the photons undergo many scattering event before leaving the sample; within the DWS formalism, their path is described as a random walk with step size $\ell^*$, the photon mean transport path~\cite{weitz_diffusing-wave_1993}. In conventional DWS, the detector is placed in the far field and the temporal evolution of the speckle pattern yields information only on the relative motion at the microscopic level, averaged over the whole illuminated sample volume. Here, by contrast, we use the Photon Correlation Imaging (PCI) configuration~\cite{duri_resolving_2009,cipelletti_simultaneous_2013}, where an image of the sample is formed. In PCI, each region of the detector corresponds to a well-defined region of the sample, so that spatially-resolved measurements of the dynamics are accessible. Additionally, any macroscopic drift of the sample entails a drift of the speckle image, so that a map of the sample strain can be obtained using cross-correlation methods similar to those applied in the analysis of the SCB images~\cite{cipelletti_simultaneous_2013}, but with no need to physically paint the sample surface.

In order to quantify both the microscopic relative motion and the macroscopic drift, the image is divided in ROIs. For each ROI, the key quantity to be computed is the two-time, spatio-temporal degree of correlation
\begin{eqnarray}
C(\zroi,\Delta y, \Delta z, t, \tau) = \\ \nonumber
\frac{\left < I(y,z,t)I(y+\Delta y,z + \Delta z,t + \tau) \right >_{\zroi}}
{\left < I(y,z,t) \right >_{\zroi}\left < I(y+\Delta y,z + \Delta z,t + \tau) \right >_{\zroi}} -1\,,
\label{eq:degcorr}
\end{eqnarray}
where $\zroi$ is the $z$ coordinate of the center of the ROI, which has the same width as the sample width and height 9.35 mm, $I(y,z,t)$ is the intensity (corrected for the dark background {\color {myc} and the uneven spatial distribution of the incident beam intensity as detailed in Ref.}~\cite{duri_time-resolved-correlation_2005}) of a pixel of coordinates $(y,z)$ at time $t$, $\Delta y$ and $\Delta z$ are spatial shifts, $\tau$ a time lag, and $\left < \cdot \cdot \cdot \right >_{\zroi}$ indicates an average over all pixels belonging to the ROI. For the sake of clarity, we first discuss the behavior of $C$ in the absence of drift motion, we then switch to the case of a pure translation (no microscopic dynamics), and finally consider the general case where both microscopic dynamics and drift coexist.

The first case corresponds essentially to traditional DWS, as reviewed, e.g., in~\cite{weitz_diffusing-wave_1993}. In the absence of drift, one sets $\Delta  y = \Delta z = 0$ in Eq.~(\ref{eq:degcorr}) and studies the two-time intensity autocorrelation function
\begin{equation}
g_2(\zroi,t,\tau)-1 = \beta^{-1}\left <C(\zroi,0,0,t',\tau) \right >_{t \le t' \le t+T}\,,
\label{eq:g2m1}
\end{equation}
where $C$ is averaged over a short time window to improve the statistics ($T = 10$ s in this work) and $\beta \lesssim 1$ is a setup-dependent constant~\cite{berne_dynamic_1976} chosen such that $g_2-1 \rightarrow 1$ for $\tau \rightarrow 0$. The intensity autocorrelation function is directly related to the mean squared displacement $<\Delta r^2(t,\tau)>$ of the scatterers between time $t$ and $t+\tau$: in the backscattering geometry of our experiments, one has~\cite{weitz_diffusing-wave_1993}
\begin{equation}
g_2(\zroi,t,\tau)-1 = A \exp\left[ -2\gamma \sqrt{k_0^2 <\Delta r^2(t,\tau)>+ \frac{3\ell^*}{\ell_a}}\right]\,,
\label{eq:g2m1dr2}
\end{equation}
with $k_0 = 2\pi n \lambda^{-1}$ the wave vector of the incident light, $n$ the sample refractive index ($n = 1.68$ for PEEK), and $\gamma$ a numerical coefficient that depends on the polarization of the incident and detected light. For a polarized incident beam, as in our experiments, Ref.~\cite{mackintosh_polarization_1989} quotes $\gamma = 1.5$ (respectively, $\gamma = 2.7$) when a polarizer is used to detect only the light with polarization parallel (respectively, perpendicular) to the $z$ axis. In the experiments described below, no polarizer is placed in front of the detector; we thus use $\gamma = 1.8$, an average of the two values of Ref.~\cite{mackintosh_polarization_1989} weighted by the relative intensity of the two polarization components of the backscattered light. The term $3\ell^*/\ell_a$ accounts for light absorption~\cite{weitz_diffusing-wave_1993}, with $\ell_a$ the absorption length, i.e. the length over which the intensity of light propagating through the sample drops by a factor $e^{-1}$. The normalization coefficient $A = \exp \left (2\gamma\sqrt{3\ell^*/\ell_a} \right)$ insures that $g_2(0)-1 = 1$.

The temporal dependence of $<\Delta r^2>$ is ruled by the nature of the microscopic dynamics. During the traction phases, the sample undergoes an affine deformation and, possibly, additional non-affine microscopic displacements. Neglecting the drift associated with the average displacement of the scatterers and assuming that thermal motion is negligible, the mean square displacement for a sample undergoing affine deformation is~\cite{bicout_diffusing_1993,crassous_diffusive_2007,erpelding_diffusive_2008,erpelding_diffusing-wave_2013}
\begin{equation}
<\Delta r^2(\tau)> =  3\ell^{*2} f\left[\mathbf{U}(\tau)\right]\,,
\label{eq:g2affine}
\end{equation}
where $f(\mathbf{U}) = [\mathrm{Tr}^2(\mathbf{U}) + 2\mathrm{Tr}(\mathbf{U}^2)]/15$ is a function of the strain tensor $\mathbf{U}$. For a tensile strain in the $z$ direction of constant rate $\dot{\varepsilon}$ applied to a material with Poisson's ratio $\nu$, one has
\begin{equation}
\label{eq:U}
\mathbf{U}(\tau) = \begin{bmatrix}
-\nu \dot{\varepsilon} \tau & 0 & 0 \\
0 & -\nu \dot{\varepsilon} \tau  & 0 \\
0 & 0 & \dot{\varepsilon} \tau  \end{bmatrix}
\end{equation}
and hence
\begin{equation}
f\left[\mathbf{U}(\tau)\right] = \frac{(\dot{\varepsilon} \tau)^2}{15}\left[(1-2\nu)^2 + 2(1+2\nu^2) \right]\,.
\label{eq:fU}
\end{equation}
As shown by Eqs.~\ref{eq:g2m1dr2} and~\ref{eq:g2affine}, the decay rate of $g_2-1$ depends on $ \ell^*$. This is because in DWS photon paths separated by more than $\ell^*$ are totally uncorrelated and do not contribute to $g_2-1$. Thus, the decay of $g_2-1$
is essentially ruled by $\ell^* \dot{\varepsilon}$, the speed at which two scatterers initially separated by a distance $\ell^*$ move with respect to each other, due to the imposed strain. 

If the sample undergoes a pure translation, the spatio-temporal correlation coefficient $C$ exhibits a peak at a spatial lag $(\Delta y , \Delta z)$ corresponding to the rigid shift between times $t$ and $t + \tau$. As long as no microscopic dynamics occur, the height of the peak is always the same, regardless of the magnitude of the displacement. By locating the position of the peak, the sample displacement can be obtained, similarly to the image analysis by stereo-DIC. In order to obtain the displacement with sub-pixel resolution, we use a dedicated algorithm that is optimized for the case where the characteristic size of the pattern (i.e. the speckle size) is of the order of the pixel size~\cite{cipelletti_simultaneous_2013}. The typical uncertainty on the displacement between two images is about 0.05 pixels, corresponding to $2.5\um$.

In the general case, the sample both translates and undergoes some microscopic dynamics. This is the case of our tensile tests, where any given ROI is shifted in the $z$ direction while being stretched. Thus, the speckle image both translates (due to the sample drift) and changes (due to the microscopic dynamics associated with the affine deformation and any additional plastic rearrangement). Accordingly, $C$ exhibits a peak, whose position shifts in time and whose height decreases as $\tau$ increases. As discussed in~\cite{cipelletti_simultaneous_2013}, in order to measure the microscopic dynamics one should in principle measure the height of the peak of $C$, rather than simply evaluating the correlation coefficient in $\Delta y = \Delta z=0$, as in Eq.~(\ref{eq:g2m1}). However, we find no significant difference between the two methods, because the peak height decreases much more rapidly than its position shifts. This is a consequence of the extreme sensitivity of DWS to the microscopic motion of the scatterers: from Eq.~(\ref{eq:g2m1dr2}) one can estimate that $g_2-1$ decreases by a factor of $e$ when the mean squared displacement is as small as $2.8\times 10^{-16}~\mathrm{m}^2$. In the following, we will therefore calculate $g_2-1$ using the simpler formula Eq.~(\ref{eq:g2m1}).

\section{Results and discussion}
\label{sec:results}

\begin{figure}
\includegraphics[width=0.9\columnwidth,clip]{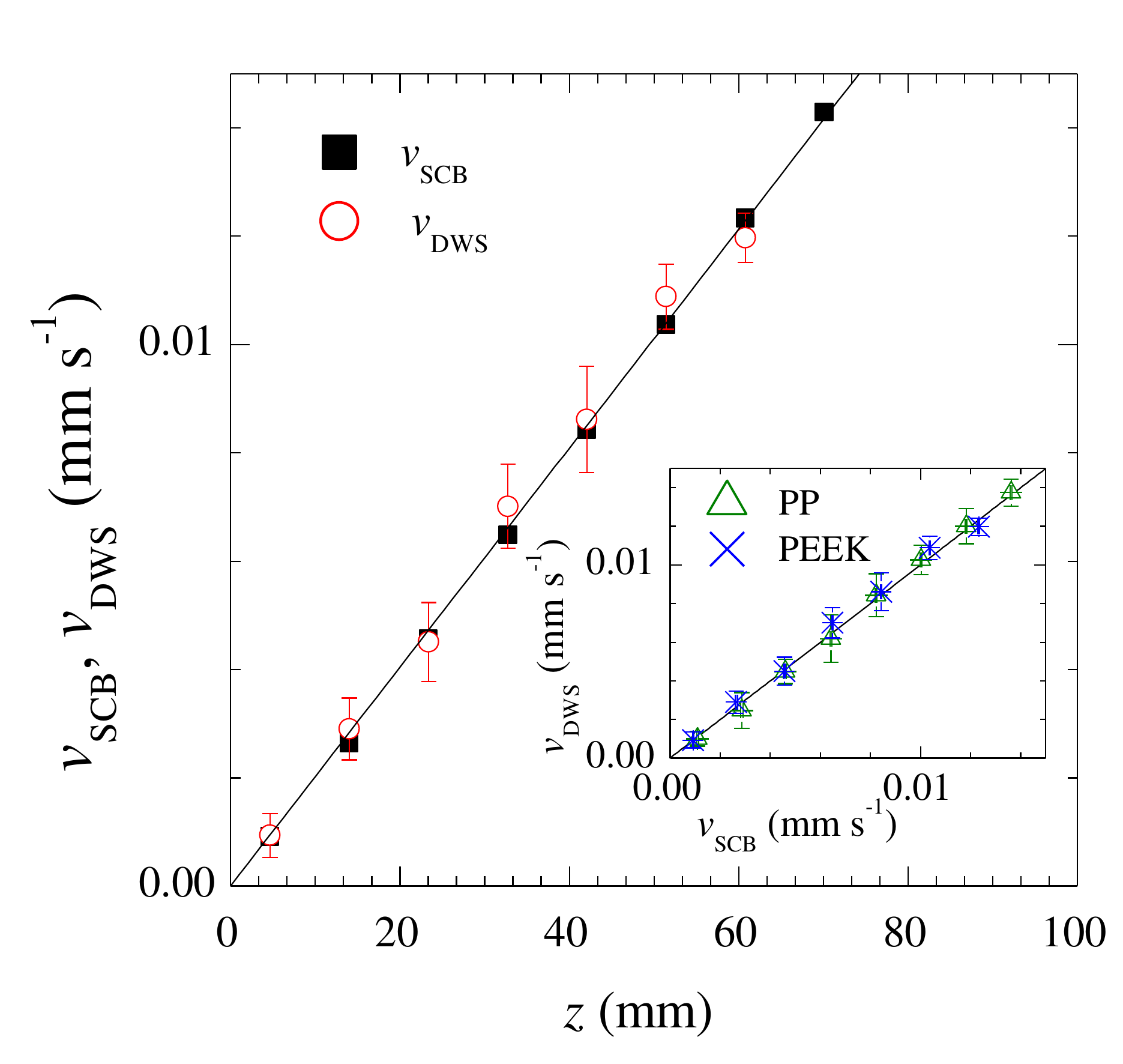}
\caption{\textbf{Comparison of the velocity profiles.} Main figure: velocity profiles during the first stretching phase as measured by conventional stereo-DIC ($\vscb$, black squares) and by space-resolved DWS ($\vdws$, open red circles), for a PEEK sample. The line indicates the imposed velocity profile, assuming a perfectly affine deformation. Inset: $\vdws$ \textit{vs} $\vscb$ for the same data as in the main panel (blue crosses) and for a PPG sample (open green triangles).}
\label{fig:vcomp}
\end{figure}
We  compare in fig.~\ref{fig:vcomp} the velocity profile during the first traction phase, as measured by stereo-DIC and by PCI-DWS. The stereo-DIC velocity profile, $\vscb(z)$, is calculated as the displacement between two consecutive images, divided by the time lag $\tau = 10$ s. As seen in fig.~\ref{fig:vcomp} (solid squares), $\vscb(z)$ increases linearly with the distance $z$ from the fixed grip, as expected for a purely affine deformation. Indeed, the data are in excellent agreement with $v_{aff}$, the theoretical profile calculated assuming affine deformation (line in fig.~\ref{fig:vcomp}). The error bars, calculated as the standard deviation over the measured displacement for the ROIs at the same $z$ but different coordinate $y$ along the width of the sample, are smaller than the symbol size. The velocity profile obtained by PCI-DWS ($\vdws$, open circles) is in very good agreement with both the theoretical expectations and the stereo-DIC one. To obtain $\vdws$, we measure the displacement $\Delta z_{\mathrm{DWS}}(\zroi,t,\tau)$ of a ROI located at $z=\zroi$ between times $t$ and $t+\tau$. The $\tau$ dependence of $\Delta z_{\mathrm{DWS}}$ is then fitted by a straight line through the origin, whose slope yields the local velocity between times $t$ and $t+\tau$. $\vdws$ and its uncertainty are finally obtained as the average and standard deviation of the instantaneous velocity thus calculated, for $t$ in the same 10-second interval as for the pair of SCB images. The choice of the maximum $\tau$ for which the linear fit is performed is optimized for each ROI: if $\tau$ is too small, the displacements are close to the measurement uncertainty, resulting in a poor determination of $\vdws$. Conversely, if $\tau$ is too large the speckle pattern is not only shifted, but it is also significantly shuffled, because of the relative motion of the scatterers associated with affine deformation, as discussed in Sec.~\ref{sec:SRDWS}. This makes it impossible to detect the displacement in a reliable way using the cross-correlation methods sketched above. For the data of fig.~\ref{fig:vcomp}, typical values of the maximum delay used for determining $\vdws$ range from 2 to 5 s. We emphasize that this correspond to displacements as small as 0.1 pixels, which highlights the importance of measuring the speckle motion with sub-pixel resolution. Finally, we note that the error bars on $\vdws$ are of order $6 \times 10^{-4}~\mathrm{mm~s^{-1}}$. This correspond to a relative error of about 10\% for an intermediate ROI, about a factor of 20 larger than the uncertainty on $\vscb$. Part of this uncertainty may be due to slight fluctuations of the pulling velocity during the test, since $\vdws$ is calculated by analyzing the 100 CMOS frames available during a 10 s interval, as opposed to $\vscb$ for which only two frames are used. Moreover, in our tests the illumination conditions were not optimized for PCI-DWS, due to the need to perform simultaneous measurements with both techniques. In particular, some white light from the SCB illumination leaked onto the PCI-DWS CMOS, introducing flare that further reduced the contrast of the speckle images{\color {myc}, as seen in Figs.~\ref{fig:setup}d-e. Indeed, in our experiment the speckle contrast is $\beta  = 0.03$, a relatively low value due to the combined effect of flare, the depolarization effect of multiple scattering, light adsorption and the relatively small speckle size, $w = 1.15~\mathrm{pixel}$, with $w$ the standard deviation of a Gaussian fit to the spatial autocorrelation function of a speckle image. We emphasize however that, in spite of the non-optimal quality of the speckle images, there is an overall excellent agreement between $\vdws$ and $\vscb$, which demonstrates the robustness of the PCI-DWS method. Finally, we note that even} if the noise on $\vdws$ may be reduced by optimizing the setup, we expect it to be still larger than that achievable by conventional stereo-DIC. Indeed, both methods use cross-correlation techniques to detect the motion of a pattern; however, the painted pattern imaged by the SCB is essentially unmodified when stretching the sample, while the speckle pattern decorrelates quite rapidly, as a result of the microscopic motion associated with sample strain. Accordingly, it is more difficult to follow precisely the displacement of the speckle pattern than that of the painted surface. We note however that the very same sensitivity of DWS to the microscopic dynamics that here somehow limits the precision on $\vdws$ provides on the other hand vary valuable information on the microscopic motion and rearrangements associated with the macroscopic strain, as we shall discuss it in the following. The inset of fig.~\ref{fig:vcomp} shows $\vdws$ vs $\vscb$, for the PEEK sample (same data as in the main plot) and for a PP glass. In both cases, the data confirm that the two methods yield the same results, to within experimental uncertainties. This validates PCI-DWS as an alternative or complementary method to conventional stereo-DIC to measure macroscopic sample deformation.

\begin{figure}
\includegraphics[width=0.9\columnwidth,clip]{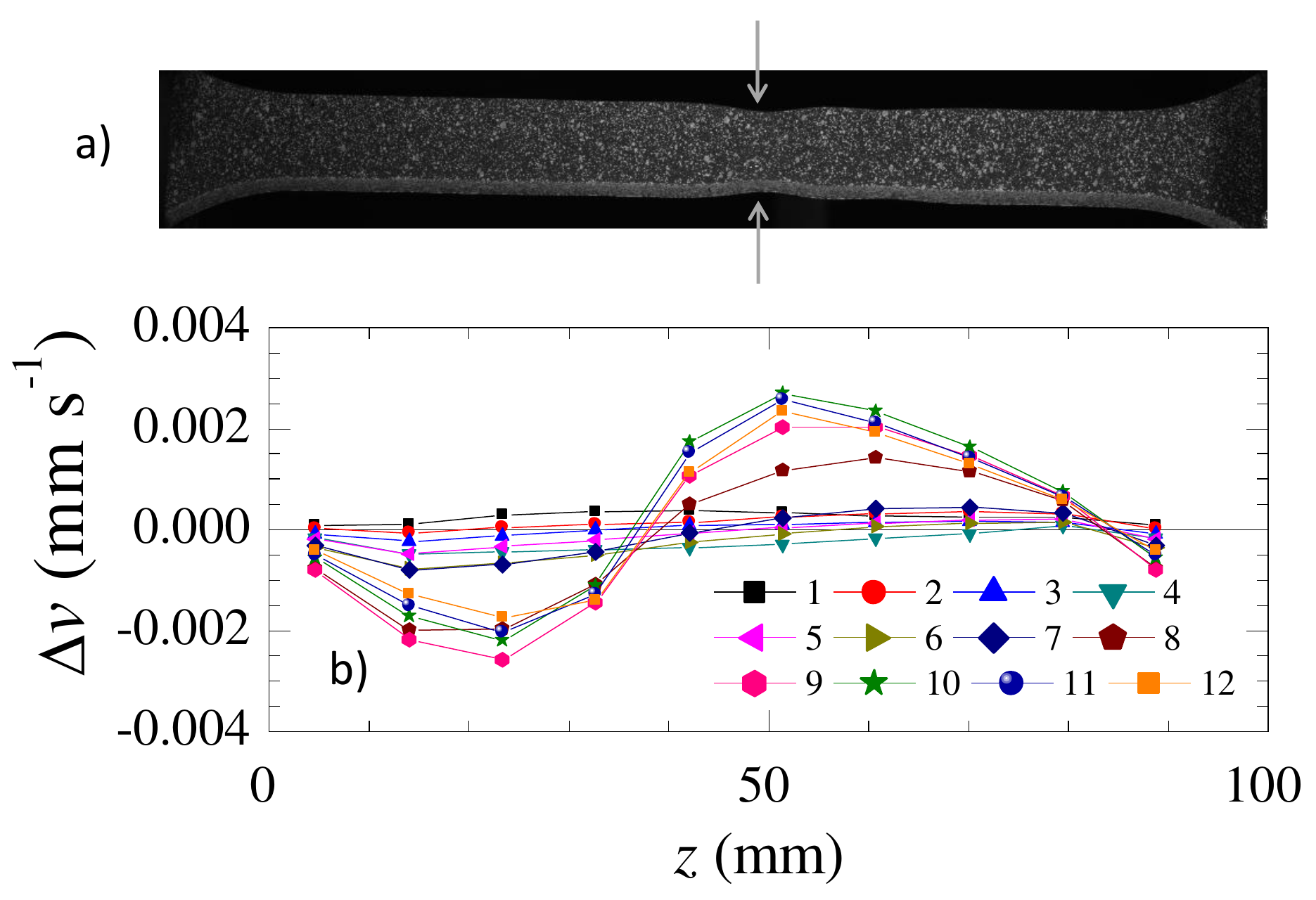}
\caption{\textbf{Onset of plastic deformation.} a): stereo correlation bench image of the PEEK sample after the 12th traction phase. A neck is clearly visible (arrows). b): deviation of the velocity profile with respect to a purely affine deformation. Data are obtained with a conventional stereo-correlation bench, for $1 \le k \le 12$ traction phases, as indicated by the labels.}
\label{fig:SCBnecking}
\end{figure}

Both the SCB and PCI-DWS images show that for a large imposed displacement of the moving grip, the strain is no more uniform, but rather concentrates around $z = 51$ mm, where the sample forms a distinctive neck, as shown in fig.~\ref{fig:SCBnecking}a. By simple visual inspection, the neck starts to be detectable after the 10-th stretching phase. We analyze both SCB and PCI-DWS data to asses whether such plastic deformation can be detected earlier by a more refined approach. Figure~\ref{fig:SCBnecking}b shows
$\Delta v = \vscb - v_{aff}$, the deviation of the SCB velocity profile with respect to affine deformation, for the 12 stretching phases. Up to the stretching phase $k =4$, $\Delta v$ is close to zero for all $z$, with no systematic behavior. For $k \ge 7$, $\Delta z$ develops a characteristic swinging appearance: close to the fix grip, the strain is less than expected ($\Delta v < 0$), while it grows rapidly with $z$ around the region where the neck will appear, and finally converges from above to the affine deformation value close to the moving grip ($\Delta v \ge 0$). For the intermediate stretching phases, $4 \lesssim k \le 6$, the behavior of $\Delta z$ is difficult to be precisely assessed. On the one hand, deviations with respect to $v_{aff}$ are comparable in magnitude to those of the first stretching phases, which may be interpreted that no deviations with respect to affine deformation are seen within experimental noise. On the other hand, a swinging behavior qualitatively similar to that seen at larger $k$ appears to develop, which might be the signature of the onset of plasticity.

\begin{figure}
\includegraphics[width=0.9\columnwidth,clip]{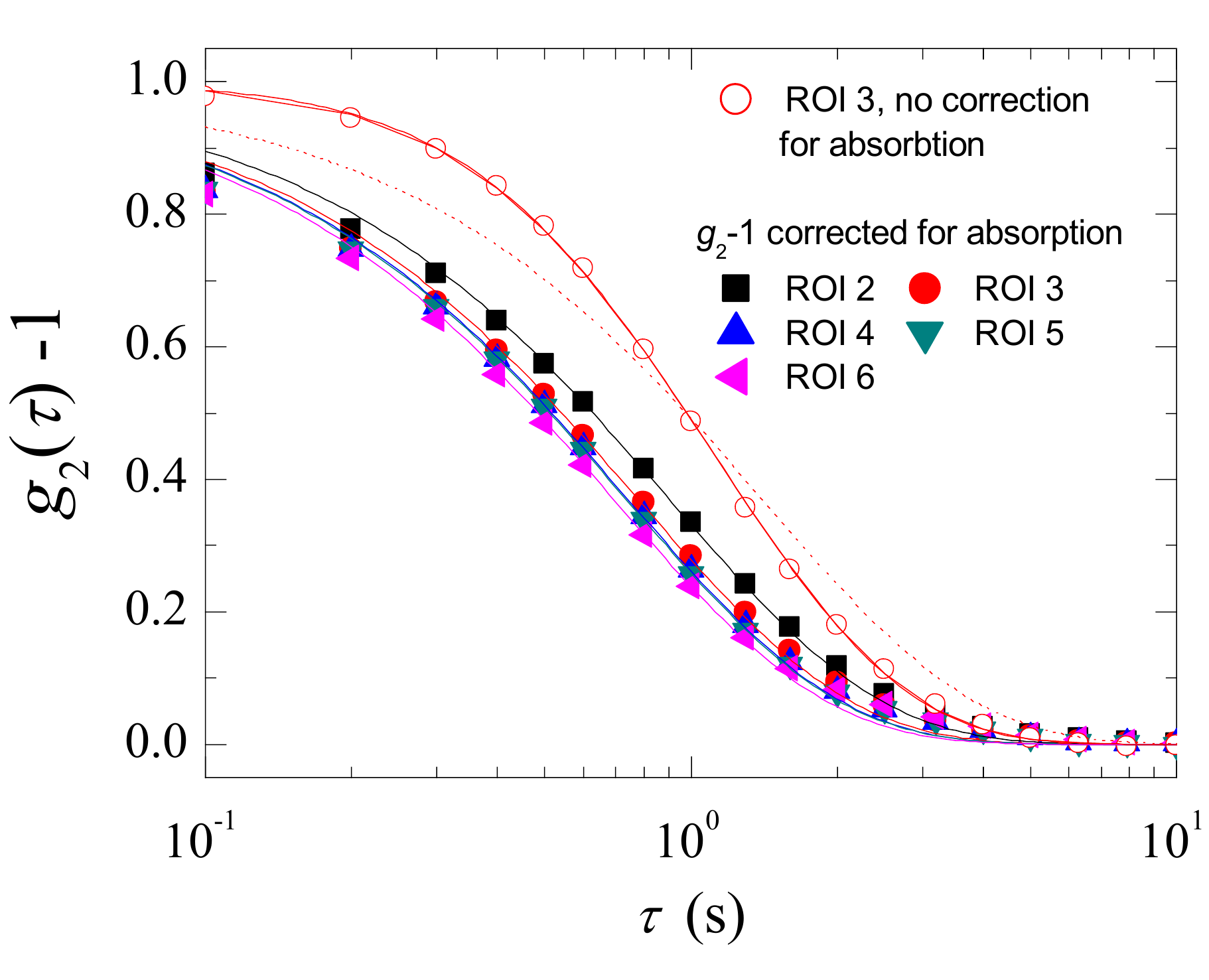}
\caption{\textbf{Microscopic dynamics during the first traction phase.} Intensity correlation function measured while applying a tensile deformation at constant strain rate. Open circles: raw data for a ROI located at $\zroi = 23.375$ mm.  The dotted (solid) red lines are the predictions for a purely affine deformation, Eqs.~\ref{eq:g2m1dr2}-\ref{eq:fU}, where the effect of absorption has not (has) been taken into account. Solid symbols: $g_2-1$ corrected for the effect of absorbtion for ROIs with $14.025~\mathrm{mm} \le \zroi = \le 51.425~\mathrm{mm}$. The lines are fits for a purely affine deformation, yielding $\dot{\varepsilon} = (5.0 \pm 0.5) \times 10^{-4}~\mathrm{s}^{-1}$.}
\label{fig:DWSnecking1}
\end{figure}

The deviations of the stretching speed with respect to an affine deformation shown in fig.~\ref{fig:SCBnecking} are at most of a few $10^{-3}~\mathrm{mm~s^{-1}}$, comparable to the typical uncertainty on $\vdws$. Thus, it is not possible to observe them directly using PCI-DWS. Instead, we exploit the great sensitivity of DWS to the microscopic dynamics. The open circles of Fig.~\ref{fig:DWSnecking1} show the intensity correlation function $g_2-1$ measured during the first stretching phase, for a ROI with $\zroi = 23.375~\mathrm{mm}~$. The decay of the correlation function is due to the microscopic dynamics induced by the applied strain:  we emphasize that no decay of $g_2$ is seen over at least several thousands of seconds when the sample is left unperturbed. We first attempt to fit the correlation function with the theoretical form predicted for a purely affine deformation, Eqs.~\ref{eq:g2m1dr2}-\ref{eq:fU}, assuming no absorption, $\ell^*/\ell_a = 0$. In the fit, the only adjustable parameter is the strain rate $\epsdot$. The wave vector is fixed to $k_0 = 1.66\times 10^7~\mathrm{m}^{-1}$, the Poisson ratio is set to $\nu = 0.4$~\cite{giraud_elaboration_2011}, and $\ell^* = 61.1\um$ is measured independently using the method of Refs.~\cite{haskell_boundary_1994,kienle_improved_1997,blaise_characterization_2012}. As shown by the dotted line, the fit does not capture correctly the shape of $g_2-1$. By contrast, an excellent agreement (solid red line) is obtained when including the absorption term, with a fitted value $\ell_a = 0.014~\mathrm{m}$. Similar values of $\ell_a$ are obtained by fitting the data for the other ROIs. In the following, we systematically correct $g_2-1$ for the effects of absorption using $\ell_a = (9 \pm 3)~\mathrm{mm}$, as obtained by averaging the results for all ROIs. Absorption-corrected intensity correlation functions are obtained by solving Eq.~\ref{eq:g2m1dr2} for $<\Delta r^2(\tau)>$ using the experimental $g_2-1$ and the known values of $\gamma$, $k_0$, $\ell^*$ and $\ell_a$. The corrected $g_2-1$ is then calculated from $<\Delta r^2(\tau)>$, using Eq.~\ref{eq:g2m1dr2} with $\ell^*/\ell_a = 0$. Figure~\ref{fig:DWSnecking1} shows as solid symbols the absorption-corrected $g_2-1$ for ROIs with $14.025~\mathrm{mm} \le \zroi = \le 51.425~\mathrm{mm}$. All data nearly collapse on a single curve, as expected if the decay rate is indeed controlled solely by the strain rate, which is the same throughout the sample. Moreover, the data are very well fitted by an exponential decay (solid lines), the behavior expected when the mean square displacement grows ballistically in time, as for a purely affine deformation at a constant rate. The only fit parameter is $\epsdot$: by averaging the fit results for all ROIs we find $\epsdot = (5.0 \pm 0.5) \times 10^{-4}~\mathrm{s}^{-1}$. This value is about 2.5 times larger than $\epsdot = 2.02  \times 10^{-4}~\mathrm{s}^{-1}$, the macroscopic strain rate obtained from the displacement velocity of the bottom grip. This discrepancy may stem from non-affine displacements not accounted for by Eq.~\ref{eq:U} that enhance the microscopic mobility. Indeed, numerical and experimental work on amorphous systems has highlighted the importance of non-affine motion even in the linear regime, due to local fluctuations of the elastic modulus~\cite{tanguy_continuum_2002,basu_nonaffine_2011}. Such heterogeneity of $E$ is likely to occur in PEEK, due to its complex structure, where crystalline domains coexist with amorphous regions. We furthermore note that a discrepancy between the strain measured by DWS and that expected from the macroscopic material properties was also observed in Ref.~\cite{erpelding_diffusive_2008}, although with an opposite trend, since the microscopic strain was found to be about 5 times \textit{smaller} than the macroscopic one. As argued in~\cite{erpelding_diffusive_2008}, these differences may also stem from the approximations introduced in the DWS formalism (see e.g.~\cite{bicout_diffusing_1993,bicout_multiple_1994,erpelding_diffusive_2008} for more details), which result in some uncertainty on the precise value of the numerical prefactor $1/15$ in Eq.~\ref{eq:fU}. We thus conclude that DWS correctly captures the time dependence of the strain evolution and that Eqs.~\ref{eq:g2m1dr2}-\ref{eq:fU} allow one to retrieve the microscopic strain rate quantitatively, to within a factor of 2.  The DWS data taken during the first stretching phase can then be used as a benchmark against which comparing the measurements for the successive stretching phases, in order to detect the onset of plasticity.

\begin{figure}
\includegraphics[width=0.9\columnwidth,clip]{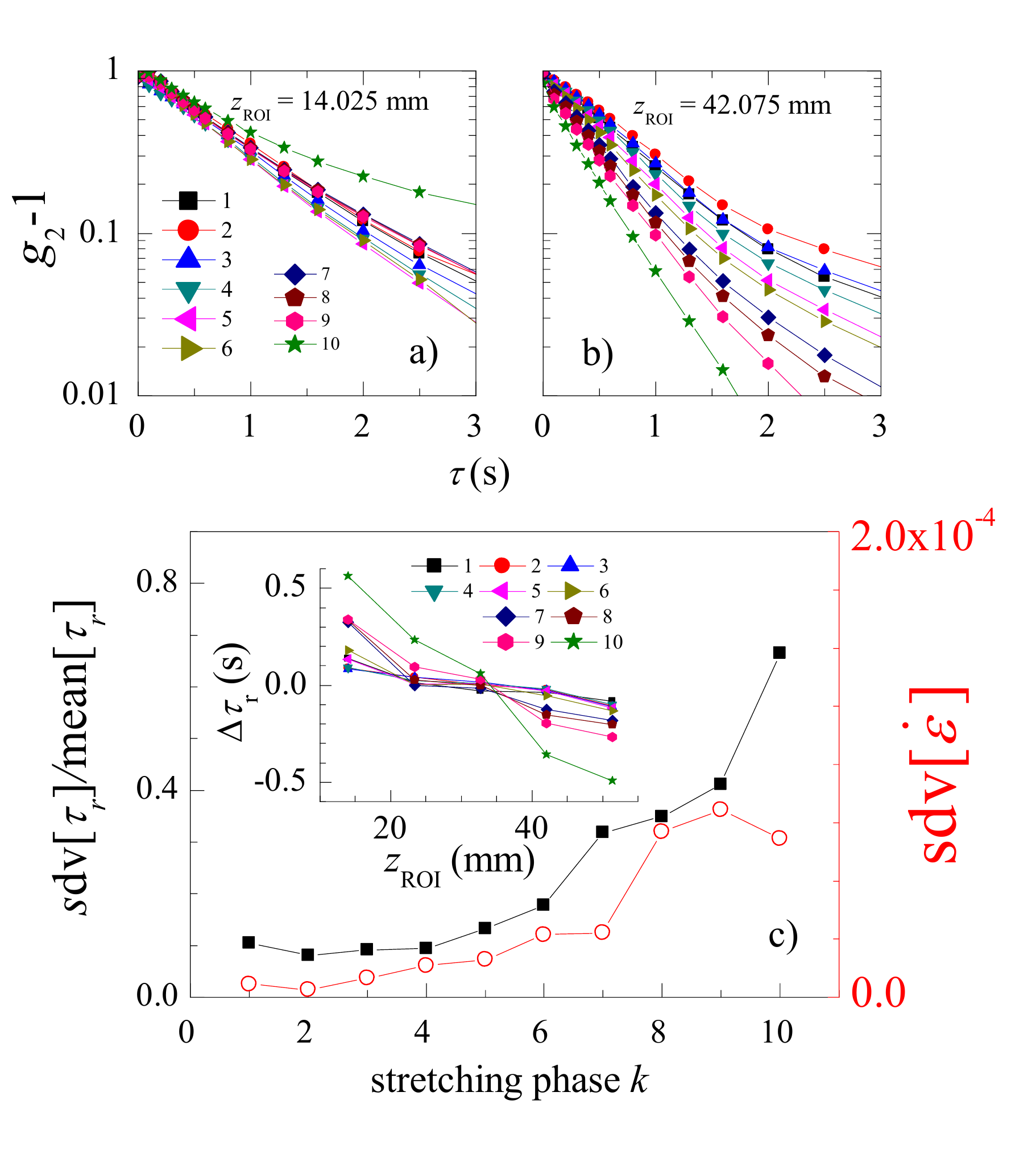}
\caption{\textbf{Microscopic dynamics and plasticity.} a) Absorption-corrected intensity correlation functions measured in a ROI close to the fixed grip, for the first 10 stretching phases, as indicated by the label. b) Same, for a ROI approximately located where a neck is macroscopically observable after the 10-th stretching phase. {\color {myc}c), inset: difference between the local relaxation time of $g_2-1$ and the spatially-averaged relaxation time as a function of ROI position. Data are labelled by the stretching phase $k$. c), main graph (left axis and solid squares): relative standard deviation of the relaxation time of $g_2-1$ measured at different locations, as a function of the stretching phase. The increase of spatial fluctuations of the relaxation time for $k\ge 5$ signals the onset of plasticity. Right axis and open circles: standard deviation of the local strain rate measured by the SCB.}}
\label{fig:DWSnecking2}
\end{figure}

We take advantage of the imaging geometry of our DWS experiments to resolve the evolution of the dynamics during the stretching phases at different sample locations. Figure~\ref{fig:DWSnecking2} shows the absorption-corrected correlation functions for various stretching phases, for two ROIs, located at $\zroi = 14.075\mm$ (a), close to the fixed grip) and $\zroi = 42.075\mm$ (b), close to where necking eventually occurs). To better appreciate the changes of the dynamics, $g_2-1$ is plotted in a semilogarithmic scale, for which the ballistic motion associated with a purely affine deformation corresponds to a straight line. Close to the fixed grip (Fig.~\ref{fig:DWSnecking2}a), the dynamics are overall consistent with ballistic motion up to the stretching phase $k = 9$. The decay becomes increasingly steeper up to $k=5$, although the change is modest. Beyond $k=5$, the decay rate follows an opposite trend: the dynamics are increasingly slower, in spite that the macroscopically imposed strain rate remains the same. For $k=10$ the dynamics strongly deviate from those expected for an affine deformation and the microscopic displacements are far more restricted than what expected from the imposed strain rate. For the ROI at $\zroi = 42.075\mm$ (Fig.~\ref{fig:DWSnecking2}b) the scenario is completely different: the dynamics speed up continuously and significantly from one stretching phase to the next one, such that the relaxation rate for $k=10$ is 2.6 times faster than that during the first stretching phase. The data shown in Fig.~\ref{fig:DWSnecking2}a,b) suggest that additional microscopic displacements, most likely due to plastic activity, enhance the dynamics since the earliest stretching phases. However, as the accumulated strain grows, these dynamics become increasingly heterogeneous: plasticity tends to concentrate in the {\color {myc} ROIs close to the region that eventually will form a neck, which in turn leads to a decrease of the dynamical activity in the other regions of the sample. This scenario is confirmed by inspecting the $z$ dependence of $\Delta \tau_r$, the difference between the local relaxation time of $g_2-1$ and its spatial average, as shown in the inset of Fig.~\ref{fig:DWSnecking2}c [$\tau_r$ is defined by $g_2(\tau_r)-1 = \exp(-1)$]. Remarkably, the changes of the microscopic dynamics and their localisation are observed since the first few stretching phases, earlier than the appearance of macroscopic heterogeneity in the strain and necking.} These observations suggest that a convenient indicator of the onset of plasticity at the microscopic level may be obtained by quantifying the spatial heterogeneity of the dynamics. To this end, we measure for each ROI the relaxation time , and plot in Fig.~\ref{fig:DWSnecking2}c) $\sigma_{rel,\tau}$, the standard deviation of $\tau_r$ over all ROIs, normalized by its mean, as a function of the stretching phase $k$ {\color {myc}(left axis and solid squares)}. The growth of this indicator for $k\ge 5$ reflects the increasing localization of the rearrangement dynamics: $\tau_r$ decreases in the region that eventually will develop a macroscopic neck, while it grows in the other ROIs. {\color {myc} For the sake of comparison, the open circles of Fig.~\ref{fig:DWSnecking2}c) show the standard deviation of the strain rate obtained from the SCB data shown in Fig. 3b. Note that the changes in the strain rate are much smaller than those of the relative relaxation time (see the difference between the scales of the left and right $y$ axes). Furthermore, a clear change of regime is seen only for $k \ge 8$, significantly later than for the DWS data. Therefore, these data demonstrate the great interest of PCI-DWS as a tool for investigating the non-linear material properties in mechanical tests.}

\section{Conclusions}
\label{sec:conclusions}
We have introduced an original setup that couples space-resolved light scattering (PCI-DWS) to  stereo-digital image correlation methods to characterize the mechanical behaviour of polymer specimens undergoing tensile loading. Light scattering allows one to measure both the mesoscopic sample displacement, yielding space-resolved strain maps, and the microscopic dynamics. The strain maps obtained by PCI-DWS are fully consistent with those issued from the stereo-correlation bench. Contrary to SCB, light scattering measurements of the displacement field do not require the surface of the sample to be spray-painted{\color {myc}, which is particularly interesting when there are concerns on the feasibility of spray coating; moreover, DWS-PCI probes the full thickness of the sample, and not only its surface, as for SCB. Thus, the method presented here is less sensitive to artifacts that may arise from a differential deformation of the surface \textit{vs.} the bulk or from a loss of stability of the spray coating, especially at large strains.} In the tests presented here, the resolution on the displacement field retrieved by PCI-DWS is somehow worse than that of SCB, although there is certainly room for improvements, e.g. by optimizing the illumination conditions. A unique feature of DWS is the possibility of quantifying the microscopic dynamics down to length scales smaller than a nm. Here, the microscopic dynamics measured by DWS during the earlier traction phases was shown to be consistent with that expected for affine deformation. By contrast, as the strain is increased, PCI-DWS reveals deviations from linear elasticity and localized plastic activity, well before any macroscopic evidence of plasticity.

{\color {myc}The requirements on the sample optical properties are those typical of DWS: as for the experiments presented here, photons should penetrate in the sample, as opposed to be reflected from its surface; $\ell^*$ should be much smaller than the sample thickness $L$, so that multiple scattering conditions are met; the absorption length $\ell_a$ should be larger than $\ell^*$, in order for the backscattered intensity to be high enough to be conveniently detected. Many materials such as semicrystalline or amorphous polymers and elastomers meet these criteria. For other materials, the method proposed here could also be used, with some modifications. For example, the mesoscopic strain field could be easily measured for reflecting objects such as metals, provided that their surface be rough on the micron scale, in order to backscatter light. In this case, however, no information on the microscopic bulk deformation may be retrieved. Nearly-transparent materials, for which $L \gg \ell^*$, could also be studied. In this case, the backscattered intensity would arise essentially from single scattering alone, as in Dynamic Light Scattering (DLS~\cite{berne_dynamic_1976}). Space-resolved DLS has already been applied to investigate the strain field of soft colloidal gels~\cite{brambilla_highly_2011} and biological gels~\cite{lieleg_slow_2011} under the action of gravity or internal stresses: it should be straightforward to extend it to harder materials such as plastics. Since single scattering probes microscopic motion on length scales larger than DWS, we expect DLS to be less sensitive to the microscopic rearrangements. While this is a disadvantage if one is interested in detecting the microscopic precursors of plasticity, such a lesser sensitivity allows a higher precision to be attained on the mesoscopic strain maps.

The work presented here highlights the great potential of PCI-DWS as a complementary tool to investigate the response of a material to a mechanical drive, both at the mesoscopic and microscopic level. Further developments of this method may include its application to other kinds of perturbations, such as temperature jumps or thermal cycling.}

\section{Acknowledgements}

We thank J. Crassous for helpful discussions. This work has been supported by the ANRT
under Contract No. 2014/0109 and by the ANR (FAPRES, grant n. ANR-14-CE32-0005-01).


\section*{References}


\begin{thebibliography}{10}
\expandafter\ifx\csname url\endcsname\relax
  \def\url#1{\texttt{#1}}\fi
\expandafter\ifx\csname urlprefix\endcsname\relax\def\urlprefix{URL }\fi
\expandafter\ifx\csname href\endcsname\relax
  \def\href#1#2{#2} \def\path#1{#1}\fi

\bibitem{sutton_image_2009}
M.~A. Sutton, J.~J. Orteu, H.~Schreier, Image {{Correlation}} for {{Shape}},
  {{Motion}} and {{Deformation Measurements}}: {{Basic Concepts}},{{Theory}}
  and {{Applications}}, {Springer Science \& Business Media}, 2009.

\bibitem{ye_kinematic_2015}
J.~Ye, S.~Andr{\'e}, L.~Farge, Kinematic study of necking in a semi-crystalline
  polymer through {{3D Digital Image Correlation}}, International Journal of
  Solids and Structures 59 (2015) 58--72.

\bibitem{pan_twodimensional_2009}
B.~Pan, K.~Qian, H.~Xie, A.~Asundi, Two-dimensional digital image correlation
  for in-plane displacement and strain measurement: a review, Measurement
  Science and Technology 20~(6) (2009) 062001.

\bibitem{el-hajjar_adhesive_2011}
R.~F. El-Hajjar, D.~R. Petersen, Adhesive polyvinyl chloride coatings for
  quantitative strain measurement in composite materials, Composites Part B:
  Engineering 42~(7) (2011) 1929--1936.

\bibitem{grytten_use_2009}
F.~Grytten, H.~Daiyan, M.~Polanco-Loria, S.~Dumoulin, Use of digital image
  correlation to measure large-strain tensile properties of ductile
  thermoplastics, Polymer Testing 28~(6) (2009) 653--660.

\bibitem{le_cam_review_2012}
J.-B. {Le Cam}, A {{Review}} of the {{Challenges}} and {{Limitations}} of
  {{Full-Field Measurements Applied}} to {{Large Heterogeneous Deformations}}
  of {{Rubbers}}, Strain 48~(2) (2012) 174--188.

\bibitem{goodman_speckle_2007}
J.~W. Goodman, Speckle phenomena in optics: theory and applications, {Roberts
  and Company}, Englewood, 2007.

\bibitem{weitz_diffusing-wave_1993}
D.~A. Weitz, D.~J. Pine, Diffusing-wave spectroscopy, in: W.~Brown (Ed.),
  Dynamic {{Light}} scattering, {Clarendon Press}, Oxford, 1993, pp. 652--720.

\bibitem{wu_diffusing-wave_1990}
X.-L. Wu, D.~J. Pine, P.~M. Chaikin, J.~S. Huang, D.~A. Weitz, Diffusing-wave
  spectroscopy in a shear flow, Journal of the Optical Society of America B
  7~(1) (1990) 15.

\bibitem{bicout_diffusing_1993}
D.~Bicout, R.~Maynard, Diffusing wave spectroscopy in inhomogeneous flows,
  Physica A: Statistical Mechanics and its Applications 199~(3) (1993)
  387--411.

\bibitem{crassous_diffusive_2007}
J.~Crassous, Diffusive {{Wave Spectroscopy}} of a random close packing of
  spheres, The European Physical Journal E 23~(2) (2007) 145--152.

\bibitem{erpelding_diffusive_2008}
M.~Erpelding, A.~Amon, J.~Crassous, Diffusive wave spectroscopy applied to the
  spatially resolved deformation of a solid, Phys. Rev. E 78~(4) (2008) 046104.

\bibitem{erpelding_diffusing-wave_2013}
M.~Erpelding, B.~Dollet, A.~Faisant, J.~Crassous, A.~Amon, Diffusing-{{Wave
  Spectroscopy Contribution}} to {{Strain Analysis}}: {{Diffusing-Wave
  Spectroscopy Contribution}} to {{Strain Analysis}}, Strain 49~(2) (2013)
  167--174.

\bibitem{wiederseiner_refractive-index_2011}
S.~Wiederseiner, N.~Andreini, G.~Epely-Chauvin, C.~Ancey, Refractive-index and
  density matching in concentrated particle suspensions: a review, Experiments
  in Fluids 50~(5) (2011) 1183--1206.

\bibitem{etcheverry_glass_2012}
M.~Etcheverry, S.~E. Barbosa, Glass {{Fiber Reinforced Polypropylene Mechanical
  Properties Enhancement}} by {{Adhesion Improvement}}, Materials 5~(12) (2012)
  1084--1113.

\bibitem{duri_resolving_2009}
A.~Duri, D.~A. Sessoms, V.~Trappe, L.~Cipelletti, Resolving {{Long-Range
  Spatial Correlations}} in {{Jammed Colloidal Systems Using Photon Correlation
  Imaging}}, Phys. Rev. Lett. 102 (2009) 085702--4.

\bibitem{cipelletti_simultaneous_2013}
L.~Cipelletti, G.~Brambilla, S.~Maccarrone, S.~Caroff, Simultaneous measurement
  of the microscopic dynamics and the mesoscopic displacement field in soft
  systems by speckle imaging, Optics Express 21 (2013) 22353--22366.

\bibitem{duri_time-resolved-correlation_2005}
A.~Duri, H.~Bissig, V.~Trappe, L.~Cipelletti, Time-resolved-correlation
  measurements of temporally heterogeneous dynamics, Phys. Rev. E 72
  (2005/11/00/) 051401--17.

\bibitem{berne_dynamic_1976}
B.~J. Berne, R.~Pecora, Dynamic {{Light Scattering}}, {Wiley}, New York, 1976.

\bibitem{mackintosh_polarization_1989}
F.~C. MacKintosh, J.~X. Zhu, D.~J. Pine, D.~A. Weitz, Polarization memory of
  multiply scattered light, Phys. Rev. B 40~(13) (1989) 9342--9345.

\bibitem{giraud_elaboration_2011}
I.~Giraud, Elaboration d'ensimages thermoplastiques thermostables:
  {{Influence}} sur le comportement m{\'e}canique des composites {{PEEK}} /
  {{Fibres}} de carbone, Ph.D. thesis, Universit{\'e} Toulouse III - Paul
  Sabatier, Toulouse (Jul. 2011).

\bibitem{haskell_boundary_1994}
R.~C. Haskell, L.~O. Svaasand, T.-T. Tsay, T.-C. Feng, M.~S. McAdams, B.~J.
  Tromberg, Boundary conditions for the diffusion equation in radiative
  transfer, Journal of the Optical Society of America A 11~(10) (1994) 2727.

\bibitem{kienle_improved_1997}
A.~Kienle, M.~S. Patterson, Improved solutions of the steady-state and the
  time-resolved diffusion equations for reflectance from a semi-infinite turbid
  medium, Journal of the Optical Society of America A 14~(1) (1997) 246.

\bibitem{blaise_characterization_2012}
A.~Blaise, C.~Baravian, J.~Dillet, L.~Michot, S.~Andr{\'e}, Characterization of
  the mesostructure of {{HDPE}} under ``in situ'' uniaxial tensile test by
  incoherent polarized steady-light transport, J. Polym. Sci. B Polym. Phys.
  50~(5) (2012) 328--337.

\bibitem{tanguy_continuum_2002}
A.~Tanguy, J.~P. Wittmer, F.~Leonforte, J.-L. Barrat, Continuum limit of
  amorphous elastic bodies: {{A}} finite-size study of low-frequency harmonic
  vibrations, Physical Review B 66~(17).

\bibitem{basu_nonaffine_2011}
A.~Basu, Q.~Wen, X.~Mao, T.~C. Lubensky, P.~A. Janmey, A.~G. Yodh, Nonaffine
  {{Displacements}} in {{Flexible Polymer Networks}}, Macromolecules 44~(6)
  (2011) 1671--1679.

\bibitem{bicout_multiple_1994}
D.~Bicout, G.~Maret, Multiple light scattering in {{Taylor-Couette}} flow,
  Physica A: Statistical Mechanics and its Applications 210~(1) (1994) 87--112.

\bibitem{brambilla_highly_2011}
G.~Brambilla, S.~Buzzaccaro, R.~Piazza, L.~Berthier, L.~Cipelletti, Highly
  {{Nonlinear Dynamics}} in a {{Slowly Sedimenting Colloidal Gel}}, Phys. Rev.
  Lett. 106 (2011) 118302.

\bibitem{lieleg_slow_2011}
O.~Lieleg, J.~Kayser, G.~Brambilla, L.~Cipelletti, A.~R. Bausch, Slow dynamics
  and internal stress relaxation in bundled cytoskeletal networks, Nature
  Materials 10 (2011) 236--242.

\end{thebibliography}

\end{document}